\documentclass[doublecol]{epl2}
\usepackage{amsmath,amsfonts,amssymb,bm,graphicx,hyperref,color,units}

\newcommand{\llangle}{\langle\!\langle}
\newcommand{\rrangle}{\rangle\!\rangle}
\newcommand{\rlangle}{\rangle\!\rangle\!\langle\!\langle}

\title{Photon-mediated electron transport in hybrid circuit-QED}

\author{Neill Lambert\inst{1}
\and Christian Flindt\inst{2}
\and Franco Nori\inst{1,3}}
\shortauthor{Lambert, Flindt \& Nori}

\institute{
  \inst{1}CEMS, RIKEN, Saitama, 351-0198, Japan\\
  \inst{2}D\'epartement de Physique Th\'eorique, Universit\'e de Gen\`eve, 1211 Gen\`eve, Switzerland\\
  \inst{3}Department of Physics, University of Michigan, Ann Arbor, MI 48109-1040, USA
}

\pacs{73.23.-b}{Electronic transport in mesoscopic systems}
\pacs{73.23.Hk}{Coulomb blockade; single-electron tunneling}
\pacs{72.70.+m}{Noise processes and phenomena}

\abstract{ We investigate photon-mediated transport processes in a hybrid circuit-QED structure consisting of two double quantum dots coupled to a common microwave cavity. Under suitable resonance conditions, electron transport in one double quantum dot is facilitated by the transport in the other dot via photon-mediated processes through the cavity. We calculate the average current in the quantum dots, the mean cavity photon occupation, and the current cross-correlations with both a full numerical simulation and a recursive perturbation scheme that allows us to include the influence of the cavity order-by-order in the couplings between the cavity and the quantum dot systems. We can then clearly identify the photon-mediated transport processes.}

\begin{document}

\maketitle

\begin{figure}
\begin{center}
\includegraphics[width=0.8\columnwidth]{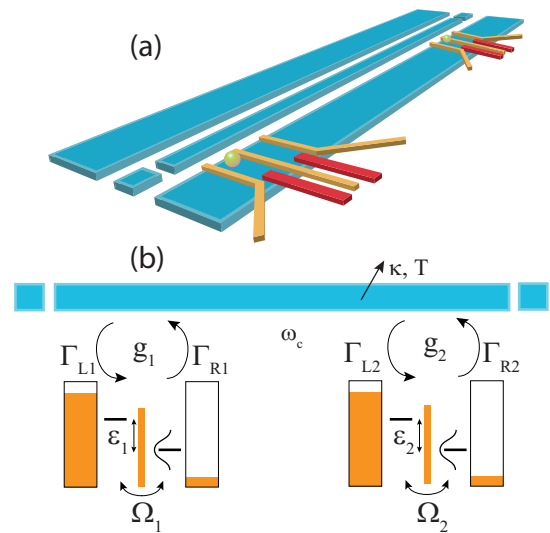}
\end{center}
\caption{(color online) Hybrid circuit-QED structure. (a) Microwave cavity capacitively coupled to separate DQDs. Voltage-biased electrodes drive currents through the DQDs. (b) The cavity is characterized by its natural frequency $\omega_c$, decay rate $\kappa$, and temperature $T$. The coupling to DQD $i$ ($=1,2$) is denoted as $g_i$. Each DQD is characterized by its level detuning $\varepsilon_i$ and tunnel coupling $\Omega_i$. Electrons enter from the left leads at rate $\Gamma_{Li}$ and leave via the right leads at rate $\Gamma_{Ri}$.}
\label{fig:schematic}
\end{figure}

\section{Introduction}

Hybrid structures that combine electronic and photonic degrees of freedom in on-chip circuit-QED architectures are currently undergoing a rapid development \cite{Xiang2013}. A series of recent experiments~\cite{Delbecq2011,Delbecq2013} have shown that controllable coupling between electronic transport in quantum dot structures and a single mode of the electromagnetic field in a microwave cavity is now achievable. Several experiments have realized both a single quantum dot and two tunnel-coupled quantum dots interacting with a microwave resonator \cite{Delbecq2011}. Very recently two quantum dots were successfully coupled to distant parts of a common cavity resonator, and photon-mediated interaction between the spatially separated quantum dot circuits was reported~\cite{Delbecq2013}.

These experimental advances are now fueling an increasing theoretical interest in understanding and predicting the physics of hybrid circuit-QED structures~\cite{Childress2004,Xiang2013}. A number of proposals~\cite{Trif2008} have already considered the coupling of electronic spins in quantum dots to microwave resonators. In parallel, other works~\cite{Jin2011} have focused on the influence of cavity resonators on the transport properties of nearby quantum dot systems. A very recent work~\cite{Bergenfeldt2013} expands theoretically on the experimental setup from Ref.~\cite{Delbecq2013} by considering two separated double quantum dots (DQDs) connected to the same cavity mode. With the DQDs weakly coupled to electronic leads at finite voltages, this system displays intriguing Tavis-Cummings physics, non-local charge transport and electronic entanglement between the DQDs. This is a promising candidate system for future experiments and further theoretical investigations are called for.

In this work we discuss a similar hybrid circuit-QED system, but focus on a different operating regime: Large voltages are now applied across the DQDs, whose electronic levels are broadened by the strong coupling to the leads, Fig.~\ref{fig:schematic}. Of particular interest is the state of the cavity induced by the out-of-equilibrium electron transport in the DQDs, as well as the photon-mediated correlations in the electronic transport.  We focus on certain resonance conditions, where the levels of one DQD are energetically detuned such that electron transport takes place via the emission of an energy quantum from the DQD to the cavity.  This cavity excitation may in turn facilitate transport in the other DQD, whose levels are oppositely detuned.

As a central goal, we examine these photon-mediated processes in the regime of weak couplings between DQDs and cavity, where we employ a recursive perturbation technique \cite{Flindt2008,Flindt2010} to calculate the average currents, the mean cavity photon occupation, as well as the cross-correlations between the currents in the DQDs \cite{Blanter2000,Lambert2007,Bergenfeldt2013}. This approach, combined with the full numerical simulation, allows us to clearly identify the photon-mediated transport processes order-by-order in the cavity couplings. Throughout the work we focus on the coupling of DQDs to a microwave cavity, but our findings are also important for other systems that couple electronic transport in DQDs to a bosonic degree of freedom \cite{Brandes2003,Ashhab2009}, for example a nano-mechanical resonator \cite{Rodrigues2005} or a vibrating molecule \cite{Flindt2005,Santamore2013}.

\section{Hamiltonian}
The system in Fig.~\ref{fig:schematic} is described by
\begin{equation}
\hat{H}=\sum_{i=1,2}\hat{H}_{e,i}+\hat{H}_{c}+\sum_{i=1,2}\hat{H}_{e{\rm -}c,i},
\end{equation}
consisting of the Hamiltonians of the electronic part ($e,i$), the cavity ($c$), and their mutual coupling ($e{\rm -}c,i$). The electronic part is made up of two DQDs that are attached to separate voltage-biased electrodes. The Hamiltonian of the individual DQD systems reads
\begin{equation}
\hat{H}_{e,i}=\hat{H}_{\mathrm{DQD},i}+\hat{H}_{\mathrm{leads},i}+\hat{H}_{T,i},
\end{equation}
where $\hat{H}_{\mathrm{DQD},i}=\varepsilon_i(\hat{n}_{Li}-\hat{n}_{Ri})/2+\Omega_i (\hat{d}_{Li}^\dag\hat{d}_{Ri}+\hat{d}_{Ri}^\dag\hat{d}_{Li})+U_i\hat{n}_{Li}\hat{n}_{Ri}$ is the Hamiltonian of DQD $i$ $(=1,2)$ with energy dealignment $\varepsilon_i$, tunnel coupling $\Omega_i$, and inter-dot Coulomb interaction $U_i$. Electrons are treated as spinless, but spin can be easily included by renormalizing the tunneling rates below. The operator $\hat{d}^\dag_{\alpha i}$ creates an electron in the left ($|\alpha i\rangle=|Li\rangle$) or right ($|\alpha i\rangle=|Ri\rangle$) level of DQD $i$ and $\hat{n}_{\alpha i}=\hat{d}^\dag_{\alpha i}\hat{d}_{\alpha i}$ is the corresponding level occupation number. The coupling to the electronic leads is described by the standard tunneling Hamiltonian $\hat{H}_{T,i}=\sum_{k,\alpha}\left[t_{k\alpha i}\hat{d}^\dag_{\alpha i}\hat{c}_{k\alpha  i}+\mathrm{h.\, c.}\right]$. The electrons in the leads are given by $\hat{H}_{\mathrm{leads},i}=\sum_{k,\alpha}\epsilon_{k\alpha i}\hat{c}^\dag_{k\alpha i}\hat{c}_{k\alpha i}$,
where $\hat{c}^\dag_{k\alpha i}$ creates an electron in the left ($\alpha=L$) or right ($\alpha=R$) lead coupled to DQD $i$. The Hamiltonian of the fundamental cavity mode with frequency $\omega_c$ is
\begin{equation}
\hat{H}_{c}=\hbar\omega_c(\hat{a}^\dag\hat{a}+1/2),
\end{equation}
where $\hat{a}^\dag$ creates a bosonic excitation in the cavity. Typically for microwave cavities this frequency lies in the range $\omega_c/2\pi \sim 5$ -- $10$ GHz. Finally, the excess charge in the DQD interacts capacitively with the cavity as
\begin{equation}
\hat{H}_{e{\rm -}c,i}=g_i(\hat{n}_{Li}-\hat{n}_{Ri})(\hat{a}^\dag+\hat{a}),
\end{equation}
where $g_i$ denotes the coupling strength. The coupling strengths $g_i$ can be related to microscopic details of the system~\cite{Childress2004}, although this will not be essential in the following. More general coupling terms can also be envisioned~\cite{Brandes2003}, but will not be considered here. The magnitude of these couplings, as estimated by recent experiments, is of the order of $g_i/2\pi \sim 50$ MHz.

\section{Generalized master equation}
To describe electron transport through the individual DQDs we trace out the electronic leads to obtain a Markovian generalized Master equation (GME) for the reduced density matrix $\hat{\rho}(t)$ of the DQDs and the cavity mode. Below we investigate the fluctuations of the electrical currents and it is useful to unravel $\hat{\rho}(t)$ with respect to the number $n_i$ of electrons that have passed through DQD $i$ during the time span $[0,t]$~\cite{Plenio1998}. Given the number-resolved reduced density matrix $\hat{\rho}(\bm{n},t)$ with $\bm{n}=(n_1,n_2)$, the joint probability of having transferred $n_{1}$ and $n_2$ electrons through DQDs $1$ and $2$, respectively, is obtained by tracing over the DQDs and the cavity,  $P(\bm{n},t)=\mathrm{Tr}\{\hat{\rho}(\bm{n},t)\}$. Moreover, associated with this probability distribution is the moment generating function $\mathcal{M}(\bm{\chi},t)=\sum_{\bm{n}}P(\bm{n},t)e^{\mathrm{i}\bm{n}\cdot\bm{\chi}}$, where $\bm{\chi}=(\chi_1,\chi_2)$ is the vector of counting fields conjugate to $\bm{n}$. Defining $\hat{\rho}(\bm{\chi},t)=\sum_{\bm{n}}\hat{\rho}(\bm{n},t)e^{\mathrm{i}\bm{n}\cdot\bm{\chi}}$, the moment generating function can be expressed as $\mathcal{M}(\bm{\chi},t)=\mathrm{Tr}\{\hat{\rho}(\bm{\chi},t)\}$. At long times, it takes the large-deviation form $\mathcal{M}(\bm{\chi},t)\approx \exp(t\Theta(\bm{\chi}))$, where $\Theta(\bm{\chi})$ is the generator of the zero-frequency cumulants of the currents. Concretely, the mean current through DQD $i$ and the corresponding zero-frequency current correlations are
\begin{equation}
I_i=e\partial_{\mathrm{i}\chi_i}\Theta(\bm{\chi})|_{\bm{\chi}\rightarrow \bm{0}},\,\, i=1,2
\end{equation}
and
\begin{equation}
S_{ij}=e^2\partial_{\mathrm{i}\chi_i}\partial_{\mathrm{i}\chi_j}\Theta(\bm{\chi})|_{\bm{\chi}\rightarrow \bm{0}},\,\, i,j=1,2.
\end{equation}
Higher-order cumulants follow through similar definitions.

We derive the GME for $\hat{\rho}(\bm{\chi},t)$ by tracing out the electronic reservoirs following Gurvitz and Prager~\cite{Gurvitz1996}. The Coulomb interactions on the DQDs are strong, $U_i\rightarrow\infty$, such that neither of them can be occupied by more than one electron at a time. The electronic state space for each DQD is then spanned by the empty state $|0i\rangle$ and the left and right single-particle states $|Li\rangle$ and $|Ri\rangle$. Coherences between states with different occupation numbers are excluded. Applying a large voltage across each DQD, so that the electronic levels are well inside the bias window, the GME takes the Markovian form
\begin{equation}
\frac{d}{dt}\hat{\rho}(\bm{\chi},t)=\mathcal{L}(\bm{\chi})\hat{\rho}(\bm{\chi},t)
\label{eq:GME}
\end{equation}
with the $\bm{\chi}$-dependent Liouvillean reading
\begin{equation}
\mathcal{L}(\bm{\chi})=\sum_{i=1,2}\mathcal{L}_{e,i}(\chi_i)+\mathcal{L}_{c}+\sum_{i=1,2}\mathcal{L}_{e{\rm -}c,i}.
\end{equation}
Here electron transport through the DQDs is given by
\begin{equation}
\begin{split}
\mathcal{L}_{e,i}(\chi_i)=&-\frac{\mathrm{i}}{\hbar}\left[\frac{\varepsilon_i}{2}\hat{\sigma}_{zi}+\Omega_i\hat{\sigma}_{xi},\bullet\,\right]\\
&+\Gamma_{Ri}\mathcal{D}(\hat{\sigma}_{Ri}^{\dag},\chi_i)+\Gamma_{Li}\mathcal{D}(\hat{\sigma}_{Li},0),
\label{eq:elec}
\end{split}
\end{equation}
having defined the pseudo-spin operators $\hat{\sigma}_{zi}=|Li\rangle\!\langle Li|-|Ri\rangle\!\langle Ri|$, $\hat{\sigma}_{xi}=|Li\rangle\!\langle Ri|+|Ri\rangle\!\langle Li|$, $\hat{\sigma}_{Ri}=|0i\rangle\!\langle Ri|$, and $\hat{\sigma}_{Li}=|0i\rangle\!\langle Li|$. The commutator in the first line of~(\ref{eq:elec}) corresponds to the coherent evolution of the individual DQDs, if isolated. Tunneling of electrons between leads and DQDs is described by the $\chi$-dependent Lindblad terms  $\mathcal{D}(\hat{\gamma},\chi)\hat{\rho}=e^{\mathrm{i}\chi}\hat{\gamma}^\dag\hat{\rho}\hat{\gamma}-\frac{1}{2}\{\hat{\gamma}\hat{\gamma}^\dag,\hat{\rho}\}$ in the second line. Electrons collected in the right leads are counted. Due to the large voltages, the transport is uni-directional from the left to the right leads and the temperature of the electrodes drops out of the problem. The tunneling rates $\Gamma_{\alpha i}=\Gamma_{\alpha i}(\varepsilon)=2\pi\sum_k|t_{k\alpha i}|^2\delta(\varepsilon-\epsilon_{k\alpha i})$ are constant. Below, we consider $\Gamma_{Li} \gg \Gamma_{Ri}$, which is the optimal limit for observing the effects of the electron-photon interaction.

The Liouvillean of the cavity reads
\begin{equation}
\mathcal{L}_{c}=-\frac{\mathrm{i}}{\hbar}[\hat{H}_c,\bullet\,]+\kappa \bar{n} \mathcal{D}(\hat{a},0)+\kappa (\bar{n}+1) \mathcal{D}(\hat{a}^\dag,0),
\end{equation}
where the commutator describes the coherent evolution of the cavity and we have included coupling of the cavity to an external environment at temperature $T$. The coupling strength is denoted as $\kappa$ and $\bar{n}=(1+\exp(\hbar\omega_c/k_BT))^{-1}$ is the mean equilibrium occupation of the cavity. Finally, the couplings between the cavity and the DQDs are
\begin{equation}
\mathcal{L}_{e{\rm -}c,i}=-\frac{\mathrm{i}}{\hbar}[\hat{H}_{e{\rm -}c,i},\bullet\,].
\end{equation}
In the following, we take $\hbar=1$, $k_B=1$, and $e=1$.

\section{Transport properties and cavity state}
With the GME at hand, we may investigate the transport properties of the DQDs and the state of the cavity. To this end, we formally solve the GME as $\hat{\rho}(\bm{\chi},t)=\exp(\mathcal{L}(\bm{\chi})t)\hat{\rho}^S$, where the stationary state $\hat{\rho}^S$ is defined as the normalized solution to $\mathcal{L}(\bm{0})\hat{\rho}^S=0$. At long times, the moment generating function is governed by the eigenvalue of $\mathcal{L}(\bm{\chi})$ with the largest real-part, so that $\mathcal{M}(\bm{\chi},t)=\mathrm{Tr}\{\hat{\rho}(\bm{\chi},t)\}\approx\exp(\max_j\{\lambda_j(\bm{\chi}\}t)$, where $\lambda_j(\bm{\chi})$ are the eigenvalues of $\mathcal{L}(\bm{\chi})$, and we may identify $\Theta(\bm{\chi})=\max_j\{\lambda_j(\bm{\chi})\}$ as the cumulant generating function for the currents.

Given the large state space of the hybrid system, it is a formidable task to calculate the dominant eigenvalue $\lambda_0(\bm{\chi})$ and its derivatives with respect to the counting fields. To circumvent this problem, we follow Ref.~\cite{Flindt2005} and consider the calculation of the current cumulants as a perturbation problem: We partition the Liouvillian as $\mathcal{L}(\bm{\chi})=\mathcal{L}+\mathcal{L}'(\bm{\chi})$ with $\mathcal{L}=\mathcal{L}(\bm{0})$ being the unperturbed operator and $\mathcal{L}'(\bm{\chi})=\mathcal{L}(\bm{\chi})-\mathcal{L}$ the perturbation. The dominant eigenvalue of $\mathcal{L}(\bm{0})$ is $\lambda_0(\bm{0})=0$, corresponding to the stationary state, while all other eigenvalues $\lambda_{j\neq 0}(\bm{0})$ have negative real-parts, ensuring relaxation towards the stationary state. We then calculate corrections to $\lambda_0(\bm{0})$, order-by-order in the counting fields, due to the perturbation $\mathcal{L}'(\bm{\chi})$. From the expansion of $\lambda_0(\bm{\chi})$ in the counting fields, we can identify the mean currents, the zero-frequency noise, as well as higher current cumulants. This approach yields for the mean currents~\cite{Flindt2005} $I_i=\llangle \bar{0}|\mathcal{J}_i|0\rrangle ,\,\, i=1,2$
and the zero-frequency noise
\begin{equation}
S_{ij}=\llangle \bar{0}|\mathcal{J}_i|0\rrangle\delta_{ij}-\llangle\bar{0}|(\mathcal{J}_i\mathcal{R}\mathcal{J}_j+\mathcal{J}_j\mathcal{R}\mathcal{J}_i)|0\rrangle,\,\, i,j=1,2.
\nonumber
\end{equation}
Here we make use of a bracket notation with $|0\rrangle=\hat{\rho}^S$, $\llangle\bar{0}|=\hat{1}$, and  $\llangle\bar{0}|0\rrangle=\mathrm{Tr}\{\hat{1}^\dag \hat{\rho}^S\}$ \cite{Flindt2004}. Moreover, $\mathcal{J}_i=\partial_{\mathrm{i}\chi_i}\mathcal{L}(\bm{\chi})|_{\bm{\chi}\rightarrow \bm{0}}$ are current super-operators, and $\mathcal{R}=\mathcal{Q}\mathcal{L}^{-1}\mathcal{Q}$ is the pseudo-inverse of $\mathcal{L}$ with $\mathcal{Q}=1-|0\rlangle\bar{0}|$ \cite{Flindt2005}. Finally, the average cavity occupation is $n_c=\llangle\bar{0}|\hat{a}^\dag\hat{a}|0\rrangle$.

\begin{figure}
\begin{center}
\includegraphics[width=0.73\columnwidth]{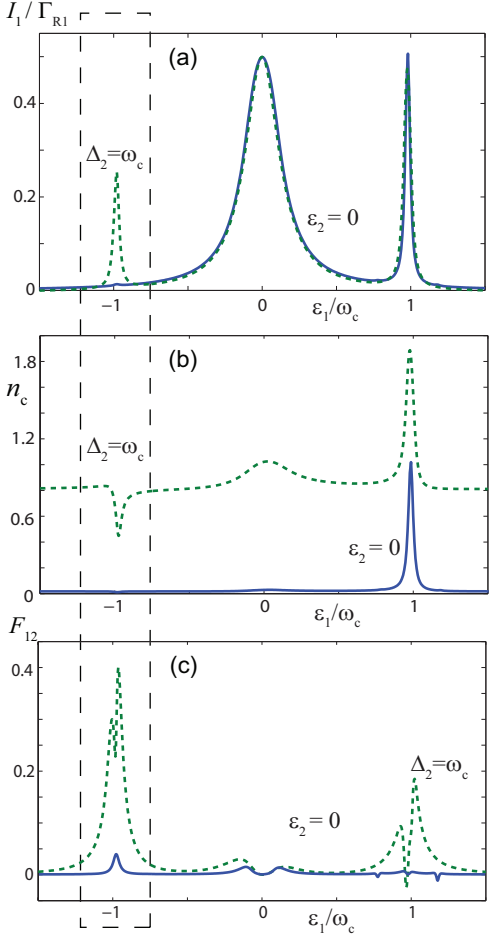}
\end{center}
\caption{(color online) Transport properties and cavity occupation. (a) Mean current $I_1$ in DQD 1. (b) Average cavity occupation $n_c$. (c) Zero-frequency cross-correlations $F_{12}=S_{12}/\sqrt{I_1I_2}$. We sweep the energy dealignment $\varepsilon_1$ in DQD 1  with $\varepsilon_2=0$ (full line) and $\Delta_2=\hbar\omega_c$ ($\varepsilon_2>0$) (dashed line) in DQD 2. Other parameters in units of the cavity frequency $\omega_c$ are $\Omega_1=\Omega_2=0.1\omega_c$, $g_1=g_2=0.05\omega_c$, $\kappa=0.005\omega_c$, $\Gamma_{L1}=\Gamma_{L2}=\omega_c$, $\Gamma_{R1}=\Gamma_{R2}=0.01\omega_c$, and $T=0$. These results come from a full non-perturbative (in $g_i$) calculation. }
\label{fig:2}
\end{figure}

In Fig.~\ref{fig:2} we sweep the energy dealignment $\varepsilon_1$ of DQD~$1$ and show, via a full non-perturbative (in $g_i$) numerical calculation, the mean current through this DQD, the average cavity occupation, and the zero-frequency cross-correlation obtained using these expressions.  The numerical calculation is cut-off for a finite number of states $N$ in the bosonic Hilbert space of the cavity mode once convergence is found. In this case the number of matrix elements in the Liouvillian scales as $O(N^4)$. Resonances occur when the natural frequencies of the DQDs, $\Delta_i=\sqrt{\varepsilon_i^2+(2\Omega_i)^2}$, match multiples of the cavity frequency, i.~e.\ $\Delta_i=m\hbar\omega_c$ for integer $m$ \cite{Brandes2003}. We consider two different dealignments of DQD 2: $\Delta_2=2\Omega_2$ ($\varepsilon_2=0$) and $\Delta_2=\hbar\omega_c$ ($\varepsilon_2>0$). The large central peak in the mean current at $\varepsilon_1=0$ is due to the levels of DQD 1 being aligned and would also occur without the cavity. The peak around $\varepsilon_1\simeq \hbar\omega_c$ corresponds to inelastic processes, where an electron tunnels from the left to the right quantum dot level of DQD 1 by emitting an energy quantum $\hbar\omega_c$ into the cavity. These processes are essentially independent of DQD 2. In contrast, the peak around $\varepsilon_1\simeq -\hbar\omega_c$ is crucially dependent on DQD 2 and only appears with $\Delta_2=\hbar\omega_c$ ($\varepsilon_2>0$). In this case, transport through DQD 2 is due to inelastic transport processes that excite the cavity mode. These excitations may in turn assist tunneling in DQD 1, where tunneling between the levels takes place via the absorption of cavity excitations.

This interpretation is supported by the average cavity occupation $n_c$ and the cross-correlated noise quantified by the Fano factor $F_{12}=S_{12}/\sqrt{I_1I_2}$ \cite{Bergenfeldt2013,Lambert2007}. With DQD 2 in resonance with the cavity, $\Delta_2=\hbar\omega_c$ ($\varepsilon_2>0$), the average cavity occupation is non-zero irrespective of DQD 1. In contrast, with $\varepsilon_2=0$ there is only an appreciable cavity occupation when DQD 1 comes into resonance at $\varepsilon_1\simeq \hbar\omega_c$. Additionally, a peak in the cross-correlated noise occurs around $\varepsilon_1\simeq-\hbar\omega_c$  when $\Delta_2=\hbar\omega_c$ ($\varepsilon_2>0$), strengthening the picture of cavity-mediated processes, where transport in DQD 1 is assisted by excitations in the cavity due to transport in DQD 2. In general, the cavity occupation and the cross-correlated noise are complicated functions of the energy dealignments $\varepsilon_i$ and thus below we focus our attention on the photon-mediated processes around $\varepsilon_1\simeq-\hbar\omega_c$, $\Delta_2=\hbar\omega_c$ ($\varepsilon_2>0$), marked with a dashed box in Fig.~\ref{fig:2}.

\section{Perturbation theory}

To single out the photon-mediated transport processes we tune the system to an operating regime where the problem can be treated using perturbation theory in the cavity couplings. Our aim is to express mean currents, average cavity occupation, and the cross-correlations as perturbative series in the cavity couplings. Concretely, we write the currents as
\begin{equation}
I_i= I_i^{(0,0)}+\frac{1}{2}\tilde{g}_1^2 I_i^{(2,0)}+\frac{1}{2}\tilde{g}_2^2 I_i^{(0,2)}+\frac{1}{4}\tilde{g}_1^2\tilde{g}_2^2 I_i^{(2,2)}+\ldots,
\nonumber
\end{equation}
where $\tilde{g}_i=g_i/\Gamma_{Ri}$, $i=1,2$, are the two dimensionless parameters that govern the expansion. Within this approach, photon-mediated processes may be identified order-by-order in the cavity couplings; e.~g.~$I_1^{(2,0)}$ is the contribution to the current in DQD 1 from a second-order process in $g_1$. We similarly write the cavity occupation and the cross-correlations as
\begin{equation}
n_c= n_c^{(0,0)}+\frac{1}{2}\tilde{g}_1^2 n_c^{(2,0)}+\frac{1}{2}\tilde{g}_2^2 n_c^{(0,2)}+\frac{1}{4}\tilde{g}_1^2\tilde{g}_2^2 n_c^{(2,2)}+\ldots
\nonumber
\end{equation}
and
\begin{equation}
S_{ij}= S_{ij}^{(0,0)}+\frac{1}{2}\tilde{g}_1^2 S_{ij}^{(2,0)}+\frac{1}{2}\tilde{g}_2^2 S_{ij}^{(0,2)}+\frac{1}{4}\tilde{g}_1^2\tilde{g}_2^2 S_{ij}^{(2,2)}+\ldots.
\nonumber
\end{equation}

To evaluate the coefficients in the expansions, we perform a new partitioning of the full Liouvillean and write it as $\mathcal{L}(\bm{\chi})=\widetilde{\mathcal{L}}+\widetilde{\mathcal{L}}'(\bm{\chi})$, where the terms that couple the DQDs to the cavity are now included in the perturbation, i.~e.~$\widetilde{\mathcal{L}}=\mathcal{L}(\bm{0})-\sum_{i=1,2}\mathcal{L}_{e{\rm -}c,i}$  and $\widetilde{\mathcal{L}}'(\bm{\chi})=\mathcal{L}(\bm{\chi})-\widetilde{\mathcal{L}}$. In this case, the unperturbed problem consists of the uncoupled cavity and DQD systems. When calculating current, noise, and higher-order current cumulants, we can then include the influence of the cavity order-by-order in the couplings to the cavity. Technically, we employ the recursive perturbation scheme developed in Refs.~\cite{Flindt2008,Flindt2010}. To calculate the mean current and the average cavity occupation we only need the corrections $|0^{(m,n)}\rrangle$ to the stationary state $|0^{(0,0)}\rrangle$ of the unperturbed problem, obeying $\widetilde{\mathcal{L}}|\bar{0}^{(0,0)}\rrangle$ and $\llangle \bar{0}|0^{(0,0)}\rrangle=1$. These corrections can be obtained from the recursive relation
\begin{equation}
|0^{(m,n)}\rrangle=-\widetilde{\mathcal{R}}\left[\frac{\mathcal{L}_{e{\rm -}c,1}}{\tilde{g}_1}|0^{(m-1,n)}\rrangle+\frac{\mathcal{L}_{e{\rm -}c,2}}{\tilde{g}_2}|0^{(m,n-1)}\rrangle  \right],
\nonumber
\end{equation}
where $\widetilde{\mathcal{R}}=\widetilde{\mathcal{Q}}\widetilde{\mathcal{L}}^{-1} \widetilde{\mathcal{Q}}$ is the pseudoinverse of $\widetilde{\mathcal{L}}$ with $\widetilde{\mathcal{Q}}=1-|0^{(0,0)}\rlangle\bar{0}|$. The coefficients entering the mean currents and the cavity occupation are then simply $I_i^{(m,n)}= \llangle\bar{0}|\mathcal{J}_i|0^{(m,n)}\rrangle$ and $n_c^{(m,n)}= \llangle\bar{0}|\hat{a}^\dag\hat{a}|0^{(m,n)}\rrangle$.
The calculation of the coefficients entering the cross-correlations is cumbersome, but essentially follows the steps described in Ref.~\cite{Flindt2010} (Sec.~IIIA) and the details are not presented here.

\section{Photon-mediated transport}

\begin{figure}
\begin{center}
\includegraphics[width=0.78\columnwidth]{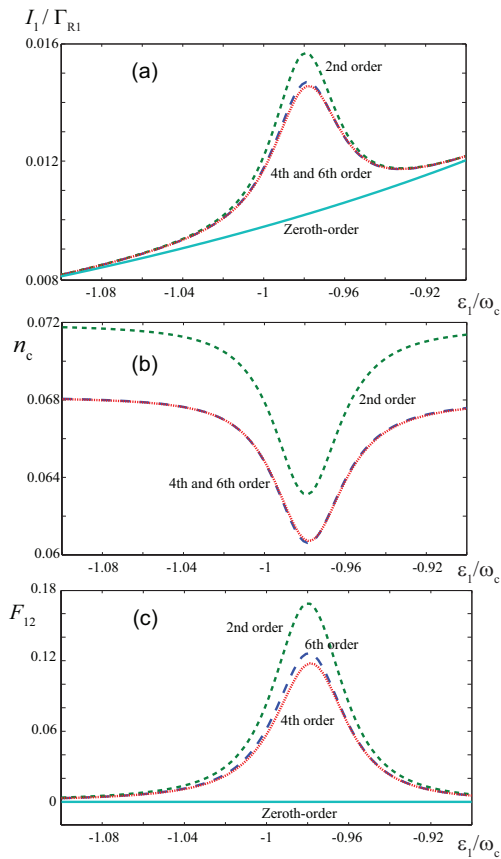}
\end{center}
\caption{(color online) Photon-mediated transport. (a) Mean current $I_1$ in DQD 1. (b) Average cavity occupation $n_c$. (c) Zero-frequency Fano factor $F_{12}=S_{12}/\sqrt{I_1I_2}$. We sweep the dealignment $\varepsilon_1$ in DQD 1 with $\Delta_2=\hbar\omega_c$ ($\varepsilon_2>0$) in DQD 2. The cavity coupling is included perturbatively, such that e.~g.~the second-order results include contributions up to second order in $g_1$ and $g_2$. For the Fano factor, we use the ratio of the noise $S_{12}$ and the currents $I_i$ up to same orders. The parameters are $\Omega_1=\Omega_2=0.1\omega_c$, $g_1=g_2=0.02\omega_c$, $\kappa=0.02\omega_c$, $\Gamma_{L1}=\Gamma_{L2}=\omega_c$, $\Gamma_{R1}=\Gamma_{R2}=0.025\omega_c$, and $T=0$.}
\label{fig:3}
\end{figure}

In Fig.~\ref{fig:3} we illustrate the use of the recursive perturbation scheme. Figure~\ref{fig:3}a shows calculations of the average current in DQD 1 with the influence of the cavity added perturbatively. The zeroth-order term $ I_1^{(0,0)}$ is the current in DQD 1 without the cavity. The second-order current is the sum of the contributions from terms proportional to $I_1^{(2,0)}$, $I_1^{(0,2)}$ and $I_1^{(2,2)}$. (Note that we consider $I_1^{(2,2)}$ as a second-order term since it is of second order in each of the couplings $\tilde{g}_1$ and $\tilde{g}_2$). The contribution from $I_1^{(2,0)}$ corresponds to a process, where a single excitation is transferred between DQD 1 and the cavity. At zero temperature, the isolated cavity is empty and it cannot assist transport in DQD 1 around $\varepsilon_1\simeq -\hbar\omega_c$, such that the contribution from $I_1^{(2,0)}$ in Fig.~\ref{fig:3}a essentially vanishes. The contribution $I_1^{(0,2)}$ is strictly zero as it only concerns the coupling between the cavity and DQD 2. The main contribution to the second-order term comes from $I_1^{(2,2)}$. This contribution is due to processes where an excitation is transferred between the two DQDs via the cavity. With $\Delta_2=\hbar\omega_c$ ($\varepsilon_2>0$), the large peak in the perturbative current, dominated by the term $\propto I_1^{(2,2)}$, around $\varepsilon_1\simeq -\hbar\omega_c$ is due to transport processes in DQD 1 facilitated by the transport in DQD 2.

A similar analysis can be carried out for the average cavity occupation and the cross-correlations. In Fig.~\ref{fig:3}b, the zeroth-order contribution $n_c^{(0,0)}$ to the cavity occupation vanishes, since the cavity at zero temperature is empty if uncoupled from the DQDs. The second-order contribution is dominated by the term proportional to $n_c^{(0,2)}$ describing cavity excitations due to tunneling in DQD 2. This contribution is independent of DQD 1 and gives a constant contribution in Fig.~\ref{fig:3}b. The second-order contribution is reduced by the term $n_c^{(2,2)}$, describing the correlated transport processes where an excitation emitted from DQD 2 is absorbed by DQD 1 instead of remaining in the cavity. The contribution from the term proportional to $n_c^{(2,0)}$ is vanishing for zero temperature.

For the cross-correlations in Fig.~\ref{fig:3}c, the zeroth-order term is strictly zero as the currents cannot be correlated without the coupling to the cavity. Similarly, the contributions from terms proportional to $S_{12}^{(2,0)}$ and $S_{12}^{(0,2)}$ are zero as they only describe the coupling of one of the DQDs to the cavity. The first non-zero contribution to the cross-correlations is contained in the term $S_{12}^{(2,2)}$, which describes the transport events where an excitation is transferred from one DQD to the other via the cavity. This contribution gives rise to the peak in Fig.~\ref{fig:3}c and is the main source of the cross-correlations. Higher-order contributions slightly reduce the cross-correlations and the series eventually converge as higher-order terms are included.

\begin{figure}
\begin{center}
\includegraphics[width=0.85\columnwidth]{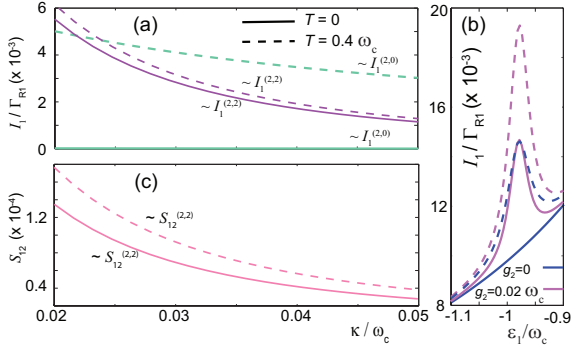}
\end{center}
\caption{(color online) Transport properties versus the decay rate $\kappa$. (a)~Current $I_1$ in DQD 1 versus $\kappa$ with two different temperatures, $T=0$ (continuous curves) and $T=0.4\omega_c$ (dashed curves). The contributions $\tilde{g}_1^2I_1^{(2,0)}/2$ and $\tilde{g}_1^2\tilde{g}_2^2I_1^{(2,2)}/4$ are shown. (b) Mean current $I_1$ in DQD 1 around the resonance $\varepsilon_1\simeq -\hbar\omega_c$ with $g_2=0$ (blue) and $g_2=0.02\omega_c$ (purple) and $\kappa=0.02\omega_c$.  (c) Cross-correlations $S_{12}$ versus the decay rate $\kappa$. The contribution $\tilde{g}_1^2\tilde{g}_2^2S_{12}^{(2,2)}/4$ is shown. The parameters are $\Delta_1=\Delta_2=\hbar\omega_c$ ($\varepsilon_1<0<\varepsilon_2$), $\Omega_1=\Omega_2=0.1\omega_c$, $g_1=g_2=0.02\omega_c$, $\Gamma_{L1}=\Gamma_{L2}=\omega_c$, and $\Gamma_{R1}=\Gamma_{R2}=0.025\omega_c$.}
\label{fig:4}
\end{figure}

To complete the picture, we finally consider individual terms in the expansions of the currents and the cross-correlations. In Fig.~\ref{fig:4}a we focus on the contributions $I_1^{(2,0)}$ and $I_1^{(2,2)}$ to the current in DQD 1 as functions of the cavity decay rate $\kappa$ for two different temperatures. We assume that DQD 2 is in resonance with the cavity, $\Delta_2=\hbar\omega_c$ ($\varepsilon_2>0$), while DQD 1 is oppositely detuned, $\Delta_1=\hbar\omega_c$ ($\varepsilon_1<0$). At zero temperature, $I_1^{(2,0)}$ is close to zero as the cavity is empty and cannot assist transport in DQD 1. The main contribution due to the cavity then comes from $I_1^{(2,2)}$, which describes the photon-mediated transport processes between DQD 1 and DQD 2.

A different situation arises with non-zero cavity temperatures. In this case, the microwave resonator is thermally occupied with photons that can assist transport in DQD 1, such that $I_1^{(2,0)}$ starts to dominate over the photon-mediated transport processes described by $I_1^{(2,2)}$. In particular, as the cavity decay rate increases, $I_1^{(2,2)}$ falls off, since photons from DQD 2 become more likely to leave the cavity instead of assisting transport in DQD 1. This makes it difficult to disentangle photon-mediated transport processes from thermally-assisted transport only by considering the average current. This is visible in Fig.~\ref{fig:4}b showing the mean current around the resonance. The results obtained for a finite temperature without DQD 2 being coupled to the cavity are very similar to the results at zero temperature including the coupling to DQD 2.

In contrast, the current cross-correlations depicted in Fig.~\ref{fig:4}c are not sensitive to the thermally-assisted transport processes, since $S_{12}^{(2,0)}$ and $S_{12}^{(0,2)}$ are strictly zero. The main contribution to the current cross-correlations comes from $S_{12}^{(2,2)}$ which directly reflects photon-mediated processes between the two DQDs. The figure shows that these processes remain even at finite temperatures and still can be identified in the current cross-correlations. Only as the cavity decay rate increases the cross-correlations reduce as photons injected from DQD 2 are increasingly likely to leave the cavity due to cavity losses.

Above, we have explicitly included cavity losses and level broadening of the electronic states due to the leads. In addition, it would be interesting to explore the influence of phonons on the photon-mediated transport processes~\cite{Brandes2005}, which may partly mask the correlated transport effects~\cite{Bergenfeldt2013}. However, future hybrid circuit-QED architectures may include carefully engineered phonon band gaps that suppress the phonon-induced dephasing rates below the strength of the DQD-cavity couplings~\cite{Weig2004}.

\section{Conclusions}

We have theoretically investigated photon-mediated transport processes in a hybrid circuit-QED architecture consisting of two separated voltage-biased DQD conductors coupled to a common microwave cavity. With both DQDs in resonance with the cavity, but oppositely detuned, photon-mediated processes take place via the cavity such that electron transport in one DQD facilitates transport in the other. By employing a recursive perturbation scheme we have evaluated the influence of the cavity on the transport properties of the DQDs order-by-order in the couplings to the microwave cavity and thereby identified the photon-mediated transport processes.

\section{Acknowledgments}
We thank C.~Bergenfeldt, J.~R.~Petta and P.~Samuelsson for useful comments. CF is supported by Swiss NSF. FN is
partially supported by the ARO, JSPS-RFBR Contract No. 12-02-92100, Grant-in-Aid for Scientific Research (S), MEXT Kakenhi on Quantum Cybernetics, and the JSPS via its FIRST Program.

\bibliographystyle{phjcp}

\end{document}